\documentclass[draft]{agujournal2019}
\usepackage{url} 
\usepackage{lineno}
\usepackage[inline]{trackchanges} 
\usepackage{soul}

\journalname{Space Weather}

\begin{document}

%
%


\title{CLEAR Space Weather Center of Excellence: All-Clear Solar Energetic Particle Prediction}

%
%




\authors{Lulu Zhao}

\affiliation{1}{Department of Climate and Space Sciences and Engineering, University of Michigan, Ann Arbor, MI, 48103, USA}




\correspondingauthor{Lulu Zhao}{zhlulu@umich.edu}


 \begin{keypoints}
 \item	The CLEAR Center will develop, test and validate a self-contained, modular (``plug and play'') framework that integrates all the major elements impacting SEPs in the inner heliosphere.

 \item	The CLEAR Center will deliver capabilities for robust and quantifiable forecasts of space radiation levels of up to 24 hours.

 \end{keypoints}

\begin{keypoints}
\item Space Weather
\item Space Radiation Prediction
\item Solar Energetic Particles
\end{keypoints}

%
%

%
%

\begin{abstract}
The CLEAR Space Weather Center of Excellence (CLEAR center) is a five year project that is funded by the NASA Space Weather Center of Excellence program. The CLEAR center will build a comprehensive prediction framework for solar energetic particles (SEPs) focusing on the timely and accurate prediction of low radiation periods (``all clear forecast") and the occurrence and characteristics of elevated periods. This will be accomplished by integrating empirical, first-principles based and machine learning (ML)-trained prediction models.
In this paper, the motivation, overview, and tools of the CLEAR center will be discussed.

\end{abstract}

\section*{Plain Language Summary}
In this paper, the Space Weather Center of Excellence, CLEAR center, will be introduced and the motivation, overview, and tools of the CLEAR center will be discussed.
The CLEAR center focuses on predicting the occurrence and properties of solar energetic particles (SEPs). In order to achieve the goal, different types of models, including empirical models, physics-based model, and machine learning models will be integrated and validated. 
The tightly-integrated team of theoreticians, observers, computational model developers, statisticians and computer scientists is uniquely qualified to accomplish this challenging task.

\section{CLEAR Center Overview}

Solar energetic particles (SEPs) can be accelerated over a wide range of energies extending up to GeVs. At relatively low energies (e.g., 10 MeV), their flux intensity can exceed the background of galactic cosmic rays by several orders of magnitude. 
The sparsity and high variability, in terms of intensity, duration, composition, and energy spectra of SEP events, the rapid arrival of the particles (often within minutes) following eruptive solar events (flares and coronal mass ejections), and our limited understanding of the conditions leading to SEP events make them difficult to predict. 

The space weather center of excellence, CLEAR, will develop, test and validate a self-contained, modular (``plug and play'') framework that integrates all the major elements impacting SEPs in the inner heliosphere: $4\pi$ maps of photospheric magnetic fields, corona ($1-20$ $R_s$), inner and middle heliosphere (0.1 AU to Jupiter's orbit), magnetic connectivity between the solar surface and any point in the inner/middle heliosphere, coronal mass ejection (CME) initiation, SEP seed population, shock acceleration and SEP transport. 
The CLEAR Center will deliver capabilities for robust and quantifiable forecasts of space radiation levels of up to 24 hours. Specifically, it will: 1) provide pre-eruption probabilistic forecasts of SEP events, 2) forecast the post-eruption SEP key parameters, including the time-dependence and maximum flux, onset, peak and end times, and the spectral characteristics, and 3) predict periods of SEP intensities below a preset threshold to issue all-clear forecasts.

\section{Motivation and Impact of the CLEAR Center}
Space radiation hazards are of great concern for aviation, satellites and astronauts. Large increases in proton flux can increase the risks of spacecraft anomalies/failures and place astronauts at risk of excessive radiation exposure.
SEPs are not only a major component of space radiation, but their intensity and energy spectra dramatically change in short periods of time. In fact, the space radiation hazard is one of the major unsolved problems (``tall poles'') hindering space travel beyond low-Earth orbit. Providing forecasts of the space radiation environment will help user communities and operational agencies to develop effective mitigation approaches. In particular, reliable all-clear forecasts are critical for, for example, maximizing opportunities for undertaking activities such as extra-vechicular activities (EVAs) or assessing the risk of SEP events occurring during spacecraft launch windows.  

Recently, NASA’s Heliophysics Division commissioned a space weather science and measurement gap analysis consistent with NASA’s role in space science and exploration. The study (hereafter the ``GAP Analysis'') \cite{Vourlidas2021} considered both Earth-based space weather users and NASA’s space exploration needs, particularly in cislunar and interplanetary space, and Mars locations. The study identified the top-level, highest priority gaps in rank order. The top ranked gap includes ``SEP occurrence and properties at a given inner heliospheric location."
Specifically, the GAP Analysis identified five SEP forecasting-related ``gaps'' that need further improvement.  While these are based heavily on operational requirements for predictions of $>\!\!10$ and $>\!\!100$~MeV proton fluxes, they also illustrate the general need to improve SEP forecasting, including ``all-clears.''
For general space mission planning and situational awareness ``All Clear'' means that the $>\!\!100$ MeV proton flux must not exceed 1 pfu. For EVAs, ``All Clear'' means the $>\!\!10$ MeV proton flux does not exceed 10 pfu with desired forecast intervals of 6h and 24h into the future. 

The CLEAR Center will focus on filling the gaps in SEP forecasting capabilities. It will deliver a new tool with readiness level (RL) of 5 that will significantly advance the top priorities identified in the GAP Analysis and provide up to 24~hr forecasts of SEP intensities, energy spectra and time profiles at any point in the inner heliosphere. We aim to improve the probabilistic forecast by increasing the probability of detection rate and reduce the false alarm rate, and improve the flux forecast by reducing the average log errors between observation and prediction with respect to the current state-of-art and the ``Space Weather Monitoring and Modeling Requirements for Beyond-LEO Missions'' released by \cite{SRAG2020}.

\section{State of Art of current SEP Modeling}
Solar flares and CMEs are responsible for accelerating energetic particles in SEP events \cite{Reames2013}.
Therefore, many currently-existing SEP models use post-eruptive observations of solar flares/CMEs to predict SEP events. \cite{Balch2008,Smart1976,Smart1989,Smart1992,Inceoglu2018,Huang2012,Belov2009,Garcia2004,Laurenza2009,Richardson2018}
There are also models make forecasts of the eruptive events (flares, CMEs, SEPs) using solar magnetic field measurements \cite{Georgoulis2008, Park2018, Bobra2016, Bobra2015, Huang2018, Boucheron2015, Falconer2014, Bloomfield2012, Colak2009,Papaioannou2015, Anastasiadis2017,Engell2017,Garcia-Rigo2016,Tiwari2015} 
In addition, because of the shorter transit times of relativistic electrons or very high energy protons compared to $\sim\!\!10$ MeV protons, near-real-time observations of $\sim$MeV electrons \cite{Posner2007} and/or $>\!\!100$MeV protons \cite{Boubrahimi2017, Nunez2015, Nunez2011} are also used to forecast the arrival of $>\!\!10$MeV protons. 
A recent review by \cite{Whitman2022} discusses more than three dozen current models to predict the occurrence probability and/or properties of SEP events. Below, we briefly summarize the three types of models that are currently in the community.

\subsection{Empirical and ML Models}
\label{subsubsec:othermodels}

These models are lightweight and typically make rapid predictions, often within seconds or minutes of the input data becoming available. For example, the SEPSTER \cite{Richardson2018} model predicts the SEP peak intensity by plugging the observed CME speed and direction into an analytical formula. Such models hold value as they can generally issue forecasts prior to the peak of an SEP event. However, since empirical and ML models are built upon historic events, it is difficult to validate their predictions at locations where no observations have been made, e.g., the journey from the earth to Mars. Furthermore, predictions can only be made for the specific energy channels upon which the model is built/trained. These models also have difficulty predicting extreme events since there are few such events available for training \cite{Bain2021, Nunez2015, Whitman2022}.

\subsection{Physics-Based Models} 

These are based on first principles \cite{Tenishev2021, schwadron2010, Alberti2017, Alho:2019, Marsh2015, Hu2017, Sokolov2004, Borovikov2015, Wijsen2020, Wijsen2022, Li2021, Luhmann2007, Aran2017, Strauss2015, Kozarev2017, Kozarev2022, Linker2019, Zhang2017}. However, physics-based models are computationally expensive and 
most physics-based models cannot make predictions based on real-time data inputs.
Moreover, nearly all physics-based models rely on observations of CMEs, and even in the best case scenario when these observations are available in near-real-time, they can barely forecast SEP events in real time. However, even if forecasts are issued after event onset, predictions of the decay phase of an SEP event may provide useful information about when conditions are expected to return to nominal.
Physics-based models are still highly attractive, since they solve for the distribution function of energetic particles and therefore they are able to provide time profiles and energy spectra of SEPs at any location of interest in the heliosphere. Moreover, ideally, if all the model parameters are known, physics-based models can predict all relevant SEP properties during all phases of the SEP event.

\section{CLEAR Center Tool Development}

The CLEAR Center will create a framework of coupled empirical, ML and physics-based SEP models that will seamlessly integrate our state-of-the-art, first-principles based, global SEP simulation capability with advanced empirical/ML methods that forecast SEP parameters. 

\subsection{Benchmark SEP Dataset}

In order to validate and evaluate the performance of the currently existing and future SEP forecast models, a standard SEP event data set is needed.
A number of SEP event lists are currently available and each spacecraft also has its own event list, e.g., the NOAA GOES proton event list, \cite{Cane2010,Richardson2014, papaioannou2016, Miteva2018b}.
However, there are discrepancies between the lists depending, for example, on the spacecraft, the particle energy examined, the event selection criteria (e.g. $>\!\!10$ MeV), and the time coverage.
In particular, lists based on the NOAA SEP data set only contain the large SEP events due to a high instrumental background as discussed further below.
Therefore, it is difficult to validate and evaluate the SEP forecast models using such incomplete and inconsistent data sets. And it is crucial that the model developers fully understand the dataset that they build the models upon, and the instruments that made the observations. 

In the CLEAR center, observers, theorists and modelers will work closely with statisticians and ML experts in constructing a reliable SEP dataset, serving as a benchmark dataset for the entire community.
We will compile a benchmark SEP event list starting from late 1973 (the launch of IMP~8) that will continue to be updated as new events are observed. The event selection criteria will be determined together by observers, theorists, statisticians, and model developers.
To identify the solar sources of the SEP events, when available, we will utilize CME, 
solar flare x-ray, visible and EUV observations of flares, and ground- and space-based solar radio emissions. 

\subsection{Foundations of Tool Development}

The University of Michigan, together with its partner institutions, will develop and deliver the groundbreaking CLEAR tool that builds on four modeling pillars that we have built and delivered to the Community Coordinated Modeling Center (CCMC):

\textbf{Space Weather Modeling Framework (SWMF)} that seamlessly couples together domain models from the solar corona to the low terrestrial atmosphere (``sun-to-mud") \cite{Toth2005, Toth2012, Gombosi2021}. The Geospace configuration of SWMF was competitively selected for operational use by NOAA/SWPC \cite{Pulkkinen2013} and has been running in operational mode since 2016 \cite{Cash2018}. The CLEAR tool will also use the SWMF and will benefit from our extensive experience in developing, validating, transitioning and supporting an operational space weather model.

\textbf{Alfv{\'en} Wave Solar-atmosphere Model (AWSoM and AWSoM-R)} \cite{Sokolov2013, vanderHolst2014a, Gombosi2018, Sokolov2021},
the cutting edge models for the time-dependent background corona/heliosphere. Any robust particle radiation capability must include a rigorous model for the background corona and heliosphere through which the particles propagate. 

\textbf{Eruptive Event Generator with Gibson-Low flux rope (EEGGL)} \cite{Gibson1998, Jin2017, Borovikov2017}, the first magnetically-driven solar-eruption model available for community use. EEGGL  enables the user to simulate the complete evolution of a CME/eruptive flare initialized with observed photospheric active-region magnetic fields.

\textbf{Multiple Field Line Advection Model for Particle Acceleration (M-FLAMPA) and Adaptive Mesh Particle Simulator (AMPS)}, two major software capabilities that model SEPs. M-FLAMPA \cite{Sokolov2004, Borovikov2018} calculates particle acceleration at shocks and propagation along a multitude of Lagrangian magnetic field lines, and AMPS \cite{Tenishev2021} is a complete 3D modeling suite for tracing particle motions. 

\subsection{Magnetic Connectivity Tool}

SEPs are high energy particles that are accelerated from close to the Sun all the way through interplanetary space. Once accelerated, these energetic particles propagate in the interplanetary plasma environment. Therefore, the magnetic connectivity between the particle source and the observation location is an important component in SEP models. In the CLEAR center, we will build a magnetic connectivity module that will be integrated into all types of models. The magnetic connectivity is often characterized by the angular distance between the particle source and the footpoint of the IMF line passing through the observer. \cite{Parker1958} spiral, potential field source surface (PFSS), and numerical MHD solutions have been used to calculate the magnetic connectivity between the source and the observer \cite{Lario2017}.

However, numerical solutions of the full set of ideal or resistive magnetohydrodynamic (MHD) equations so far have not been able to reproduce aligned interplanetary stream-lines and magnetic field lines in corotating frames.
One of the reasons for this discrepancy is the numerical reconnection across the heliospheric current sheet: the reconnected field is directed across the current sheet, while the global solar wind streams along the current sheet, thus resulting in ``V-shaped'' magnetic field lines and significant misalignment between field lines and streamlines. Within regular MHD there is no mechanism to re-establish the streamline -- field line alignment.

\begin{figure}[htb]
    \vspace{-1.5ex}
    \includegraphics[width=1\textwidth]{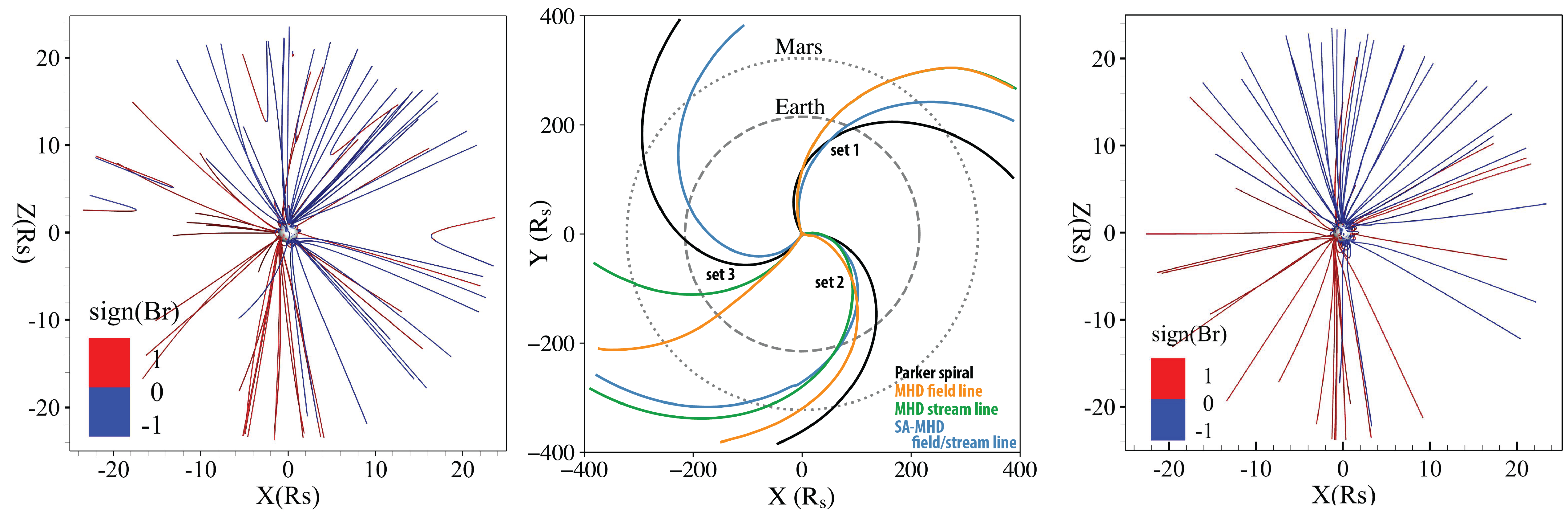}
    \caption{Left: 3D magnetic field lines in the solar corona domain ($\pm23R_s$). Red and blue color represent positive and negative magnetic polarities. ``V-shaped'' magnetic field lines form when oppositely oriented field lines reconnect (mainly across the current sheet). Middle: Three sets of field lines/stream lines in the ecliptic plane, each containing four curves: ideal \cite{Parker1958} spiral (black), MHD field line (orange), MHD stream line (green) and SA-MHD \cite[]{Sokolov2022} (blue). In set 1 the MHD field line and stream line overlap. Right: Same as the left panel, but obtained with SA-MHD \cite[]{Sokolov2022}. Note the complete absence of ``V-shaped'' magnetic field lines.
    }\label{fig:connectivity}
\end{figure}

Tracing magnetic field lines between two points of interest in the heliosphere ({\it magnetic connectivity}) is so challenging that sometimes it is preferable to trace plasma stream-lines instead \cite{Young2021}, or even Lagrangian trajectories of fluid elements. This point is illustrated in \figurename~\ref{fig:connectivity}. The MHD solution shown in \figurename~\ref{fig:connectivity} is calculated by using the Global Oscillation Network Group (GONG) magnetogram observed at 2013 Apr 11, 06:04 UT. The x-y plane is the ecliptic plane and the z-axis extends toward the poles of the sun. The left panel shows that in MHD solutions oppositely oriented magnetic field lines tend to reconnect along the heliospheric current sheet. 
This processes makes the MHD magnetic connectivity invalid (forms ``V-shaped'' field lines). In addition, the magnetic polarity of the field line ``flips'' at the top of the ``V,'' therefore the pitch angle of SEPs also flips, causing computational difficulty for physics-based transport equation solvers. 

Recently our group introduced the Stream-Aligned MHD (SA-MHD) method that ``nudges'' the magnetic field lines and plasma stream lines towards each other \cite{Sokolov2022}. This can be seen in the right panel of \figurename~\ref{fig:connectivity} which demonstrates the complete absence of ``V-shaped'' field lines in the SC domain (this feature is conserved in the entire heliosphere).

\begin{figure}[htb]
    \vspace{-2.5ex}
    \includegraphics[width=\textwidth]{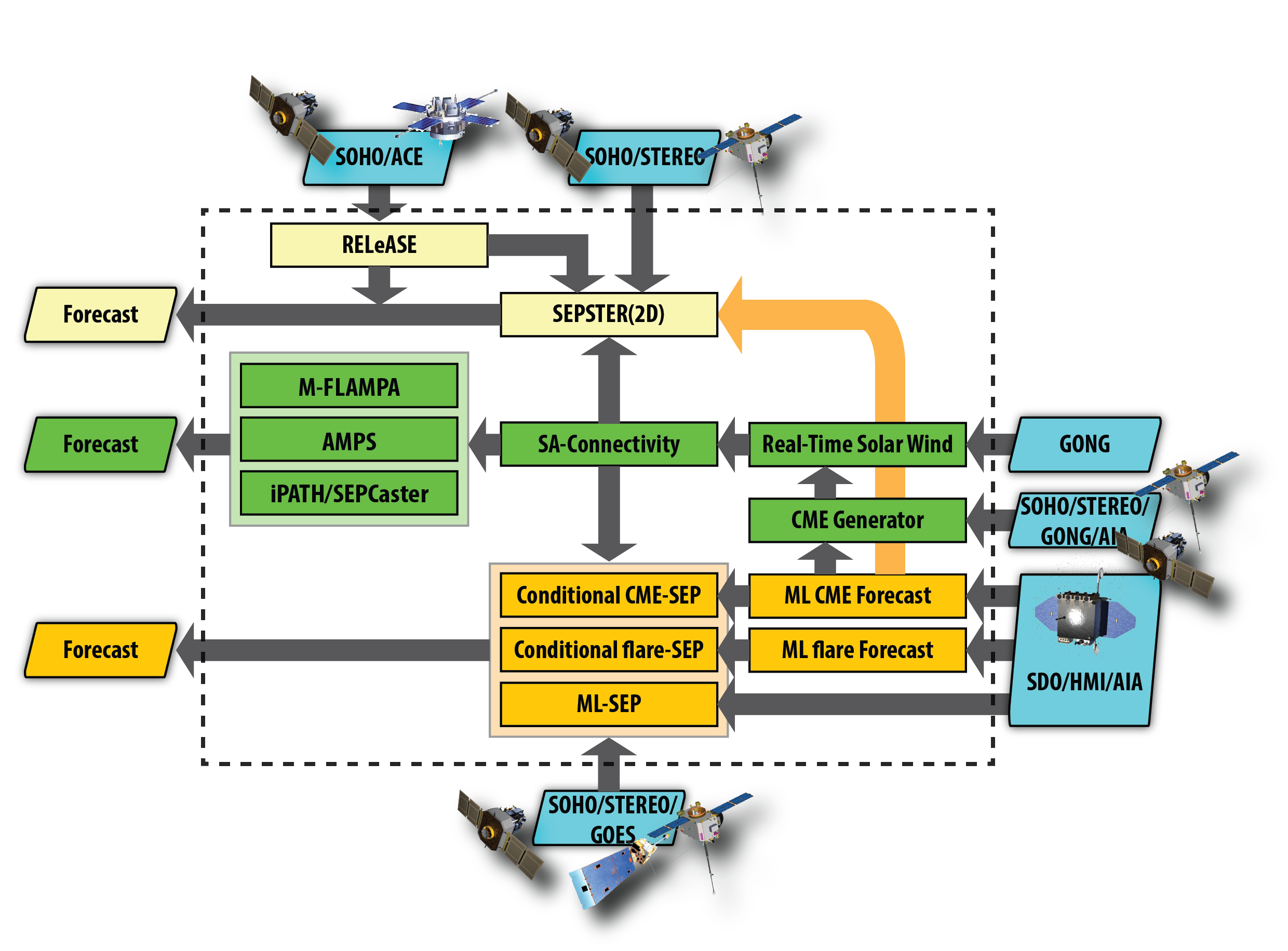}
    \caption{Overview of the CLEAR center workflow. All the observational inputs are shown in the cyan color. The empirical, ML, and physcis-based model components are shown in yellow, green, and orange, respectively.
    }\label{fig:integrate}
\end{figure}

The middle panel of \figurename~\ref{fig:connectivity} illustrates the model dependence on the magnetic connectivity. It shows various connectivity models: \cite{Parker1958} spiral, MHD field lines, MHD stream lines and SA-MHD magnetic field lines. Set 1 originates from the Sun far away from the current sheet. In this set the MHD field lines and stream lines overlap, as they are supposed to. Set 2 originates closer to the current sheet and the MHD field lines and stream lines are somewhat separated. Set 3 originates from the vicinity of the current sheet and the MHD field lines and stream lines are separated (``V-shaped'' field lines form around Mercury's orbit). More importantly, the SA-MHD field line is nearly $90^\circ$ away from the MHD field line. CLEAR will use SA-MHD to obtain magnetic connectivity and feed it to all SEP models: empirical, ML, and physics-based. This is shown in \figurename~\ref{fig:integrate}.

An additional problem with magnetic connectivity calculations is that different magnetograms can give different coronal field configurations and hence connections. While SA-MHD, to be used in this project, gives a better estimate of the connectivity for a given magnetogram.
The selection of the appropriate magnetograms is a critical element of radiation environment modeling and the CLEAR team is currently collaborating with most magnetogram providers in synergistic projects to optimize the input magnetograms.

\subsection{SEP Forecast Tools}
\subsubsection{Empirical Models}
\label{subsubsec:empirical}

\textbf{SEPSTER} (``SEP prediction derived from STEReo observations") is an empirical model built by constructing an analytical formula that relates the peak proton intensity in the energies of $14-24$ MeV to the speed and direction of the associated CME relative to the observing spacecraft's magnetic field footpoint at the Sun \cite{Richardson2018}.
A more advanced version of SEPSTER, \textbf{SEPSTER2D} \cite{Bruno2021} predicts SEP event-integrated and peak intensity proton spectra between $10$ and $130$ MeV at $1$ AU, based on cross-calibrated historical observations from GOES-13/15, STEREO~A and B, and the PAMELA mission 
\cite{Bruno2018}. 

Since SEPSTER(2D) use observations of  eruptive events, they provide now-casts of SEP events. While the model predictions can be rapidly calculated, currently the forecast lead time is limited by the lack of real time spacebased coronagraph observations. Neither the SOHO LASCO or the STEREO SECCHI coronagraphs return observations in real time, and are subject to delays, for example, due to limited contacts with ground stations that may be of the order of hours or even days; additional time is required to process and interpret the observations to derive the CME parameters. The lead time should be improved when near real time images become available from the NOAA Space Weather Follow-On L1 mission (SWFO-L1) coronagraph combined with automated CME detection. 

The CLEAR center will integrate new capabilities into SEPSTER/SEPSTER2D. First, the longitudinal and latitudinal magnetic connectivity will be calculated using the magnetic footpoint derived from SA-MHD simulations \cite{Sokolov2022}. 
Currently, the model accounts for both longitudinal and latitudinal magnetic connectivity by assuming Parker spiral field lines beyond the nominal source-surface height, $2.5$ $R_s$ \cite{Bruno2021}.
Second, to address the lack of real time CME observations, we will feed the CME parameters output by ML models, discussed below, into SEPSTER/SEPSTER2D, making them true SEP forecast models with greatly increased forecast lead time. 

\textbf{REleASE} (Relativistic Electron Alert System for Exploration) is another empirical model integrated in the CLEAR Center. It uses promptly-arriving near-relativistic or relativistic electron observations at L1 as an early indicator of an energetic proton event \cite{Posner2007}. REleASE is a nowcast model and it takes the real-time measurement of energetic electrons by ACE and SOHO, when they are available.  
A unique aspect of this forecast tool is the use of \textit{in-situ} particle measurements as opposed to reliance on remote observations of the solar eruption (e.g., solar flare X-rays). Thus, REleASE can make predictions for $\sim25$\% of SEP events with sources behind the west limb \cite{Richardson2014} which cannot be observed by ground based, Earth or L1 orbiting instruments, exemplifying the benefits of considering results from different models in the proposed framework. 
In the CLEAR center, we will evaluate the performance of the REleASE model using the benchmark dataset developed by CLEAR. We will also combine REleASE and SEPSTER/SESTER2D by requiring the detection of an electron onset by REleASE before a SEPSTER/2D prediction is issued.  This will considerably reduce the number of ``false predictions" made when SEPSTER/2D is triggered by every CME \cite{Richardson2018}. The integrations of the SEPSTER(2D) and RELeASE modules in the CLEAR center is shown in \figurename~\ref{fig:integrate}.

\subsubsection{ML Models}
\label{subsubsec:ml}
Unlike empirical models, ML models do not construct analytical formulae relating the observable parameters of solar events and SEPs. Instead, a ML model is a universal approximator, a general form depending on different ML techniques. 
However, to successfully train, validate and test the advanced ML techniques in the limited SEP database still remains a challenge. The rare event handling and the heterogeneity of the mechanism for SEP formation and identification require a lot of careful thinking on the modeling structure and data preparation, plus the validation procedure. Our team is uniquely positioned with statisticians and machine learning experts who have prior experiences in rare event handling/forecasting, and who have the capability of developing innovative, non-off-the-shelf methods that caters the need for SEP forecasting. 

Most current ML SEP prediction models are nowcast models that predict the occurrence and properties of SEPs after the onset of the solar eruptive events.
The lead times of the nowcast models are limited by the time lag between the solar eruptive events and the arrival of energetic particles, which can be as little as a few minutes. 
In order to make true forecasts of SEP events, it is required to build models that use the measurements of photospheric magnetic fields \cite{Inceoglu2018,Kasapis2022}, which are the drivers of solar eruptive events. 

There are various ways to build a ML SEP model: 1) train a model that predicts the SEP occurrence and properties using SHARP\cite{Bobra:2014a}, SMARP\cite{Bobra:2021a}, or directly using the images of the active region magnetic field measurements; 2) build a conditional SEP prediction model based on the prediction of solar eruptive events. When building the ML models, both the single frame image and the time evolution of the observational measurement will be used.

The ML components and their prediction flow in the CLEAR center are shown in \figurename~\ref{fig:integrate}. The ML model components are shown in orange.
The ML-CME model will be integrated into the physics-based model, discussed in the following section.
And when training the SEP forecast/nowcast ML models, the magnetic connectivity module as discussed above will be integrated in to the ML models.

In the CLEAR Center, the ML models will be built using SDO/HMI, SOHO/MDI, and GONG magnetic field measurements, GOES solar flare measurements, the CME parameters from SOHO/LASCO and STEREO/SECCHI, and the benchmark SEP dataset. Both SEP scientists and ML experts will be involved at every stage of the ML design, which will improve the interpretability of the ML models.

\subsubsection{Physics-based Models}
\label{subsubsec:physicsmodels}

Physics-based SEP forecast models solve equations that describe the acceleration and transport processes of energetic particles in the solar corona and interplanetary space. 
The shocks formed by fast and wide CMEs are the main contributors to particle acceleration in large gradual SEP events \cite{Kahler1994, Reames1999, Gopalswamy2008, Park2012, Kahler2013, Kahler2014}. Therefore, current physics-based models provide forecasts of gradual SEP events in which the particles are accelerated by CME-driven shocks.
A physics-based SEP forecast tool requires a real-time background solar wind module, a CME generation module, and
a particle acceleration and transport module.
It is also essential that the entire forecast system runs faster than real-time.

\subsection{Real-Time Solar Wind Tool}
\label{subsec:realtime}

Ideal Parker spiral field lines are often assumed to determine the magnetic connectivity. However, magnetic fields close to the sun ($r\!\!<\!\!20$ $R_s$) are far more complex. Beside ideal Parker spiral, the PFSS model has also been utilized to calculate the magnetic field topology close to the sun. 
The PFSS model is simple and reproduces the open fields from coronal holes well \cite{Nitta2008}, but sometimes fails to reproduce open field lines around active regions \cite{Riley2006, Nitta2008, Schrijver2011}.

\subsubsection{AWSoM/AWSoM-R}
\label{subsubsec:awsom}

In the CLEAR Center, the background solar wind plasma in which the SEPs propagate is modeled by Alfv{\'e}n Wave Solar-atmosphere Model (-Real time) (AWSoM/AWSoM-R) \cite{Gombosi2021, Gombosi2018}
driven by near-real-time hourly-updated GONG (bihourly ADAPT-GONG) magnetograms. The solar wind model has been validated by comparing simulations and observations for both the in-situ solar wind and the predicted line-of-sight (LoS) appearance of the corona in different wavelengths \cite{Sachdeva2019,Gombosi2021}. 

AWSoM \cite{vanderHolst2014a, Sokolov2021, Gombosi2021} solves the coupled MHD equations (with proton temperature anisotropy) and the transport and dissipation of Alfv{\'e}n turbulence. Excess heat dissipated in the corona is transported back (via electron heat conduction) to the chromosphere where it is lost by radiative cooling. In the transition region the plasma temperature increases some two orders of magnitude over $\sim\!\!10^2$ km. Instead of solving a computationally expensive 3D problem on a very fine grid, one can reformulate it in terms of a multitude of much simpler 1D problems along \textit{threads} that allows us to map the boundary conditions from the solar surface to the corona \cite{Sokolov2021}. It is implemented in AWSoM-R that can achieve faster than real-time performance on $\sim$512 cpu cores. 

We will utilize the newly-developed SA-MHD model, a stream-aligned version of AWSoM-R \cite{Sokolov2022} (see Figure~\ref{fig:connectivity}).
The real-time solar wind module is driven by the GONG magnetic field measurements which are publicly available at \url{https://gong.nso.edu/data/magmap/}.
The National Solar Observatory’s (NSO) provides traditional synoptic maps updated every hour, as well as standard Carrington type maps for every solar rotation. 
The methodology for assembling these photospheric field maps do not account for key dynamics occurring on the Sun such as differential rotation, meridional flow, supergranulation, and flux emergence. 
Magnetic flux transport models such as the Air Force Data Assimilative Photospheric Flux Transport (ADAPT) model \cite{Hickmann2015} will also be used to drive the solar wind model and we will select the magnetograms that best represents the observations.

\subsection{CME Generation Modules}
\label{subsec:cme}

Large SEP events are usually associated with CMEs and SEPs are believed to be accelerated by shocks driven by CMEs/ICMEs through the first-order Fermi acceleration mechanism \cite{Axford1977, Krymsky1977}. 
Although there is a good correlation between large SEP events and fast and wide CMEs, not all fast and wide CMEs have an associated SEP event \cite{Lario2020,Gopalswamy2017,Swalwell2017,Marque06}. 
Possible explanations proposed include \cite{Lario2020}: 1) CME dimension and dynamics \cite{Kahler2019a}, especially during the initial evolution phase; 2) inefficient acceleration at the shock (e.g., low Mach number); and 3) transport through the corona and interplanetary space \cite{Zhang2017, Zhao2018}.
Particles can also be continuously accelerated as ICME-driven shocks propagate out in interplanetary (IP) space, leading to so-called ``energetic storm particle'' (ESP) events, localized particle enhancements observed around shock passage. The ESP in the SEP events pose radiation hazards.
Therefore, to provide reliable forecasts of SEP events including the ESP phase, detailed simulations of the propagation of CMEs/ICMEs through the solar corona and interplanetary space are necessary.

The CME/ICME simulations in the CLEAR Center will be driven by the EEGGL (Eruptive Event Generator by  \cite{Gibson1998}) 
or the TiDeS-G (Titov-D\`emoulin-Sokolov-Gombosi) \cite{Titov1999, Sokolov2023}, which both are existing modules in the SWMF. These tools allow driving a CME in a global simulation by superimposing a \citeA{Gibson1998} (GL) or \citeA{Titov1999} (TD) magnetic flux-tube configuration onto the background state of the solar corona \cite{Manchester2004a, Manchester2004b}. These magnetic configurations describe an erupting magnetic filament. The filament becomes an expanding flux rope (magnetic cloud) in the ambient solar wind while evolving and propagating outward from the Sun, thus allowing the simulation of the propagation of a magnetically driven CME.

\subsection{Particle Acceleration and Transport Modules}
\label{subsec:acceleration}

The third module in the physics-based SEP forecast model is the particle acceleration and transport module. 
Current physics-based models adopt two approaches to modeling particle acceleration and transport processes: solving the particle distribution functions along the Lagrangian magnetic field lines, or solving the particle trajectories. 
Each approach has its advantages and limitations. In the CLEAR Center, the acceleration and transport of energetic particles are modeled by three different models: M-FLAMPA, AMPS, and iPATH/SEPCaster.
M-FLAMPA solves the distribution functions along Lagrangian magnetic field lines, and AMPS and iPATH/SEPCaster solves the trajectories of energetic particles.

\subsubsection{M-FLAMPA}
\label{subsubsec:mflampa}

\begin{figure}
    \vspace{-3ex}
    \includegraphics[width=0.45\textwidth]{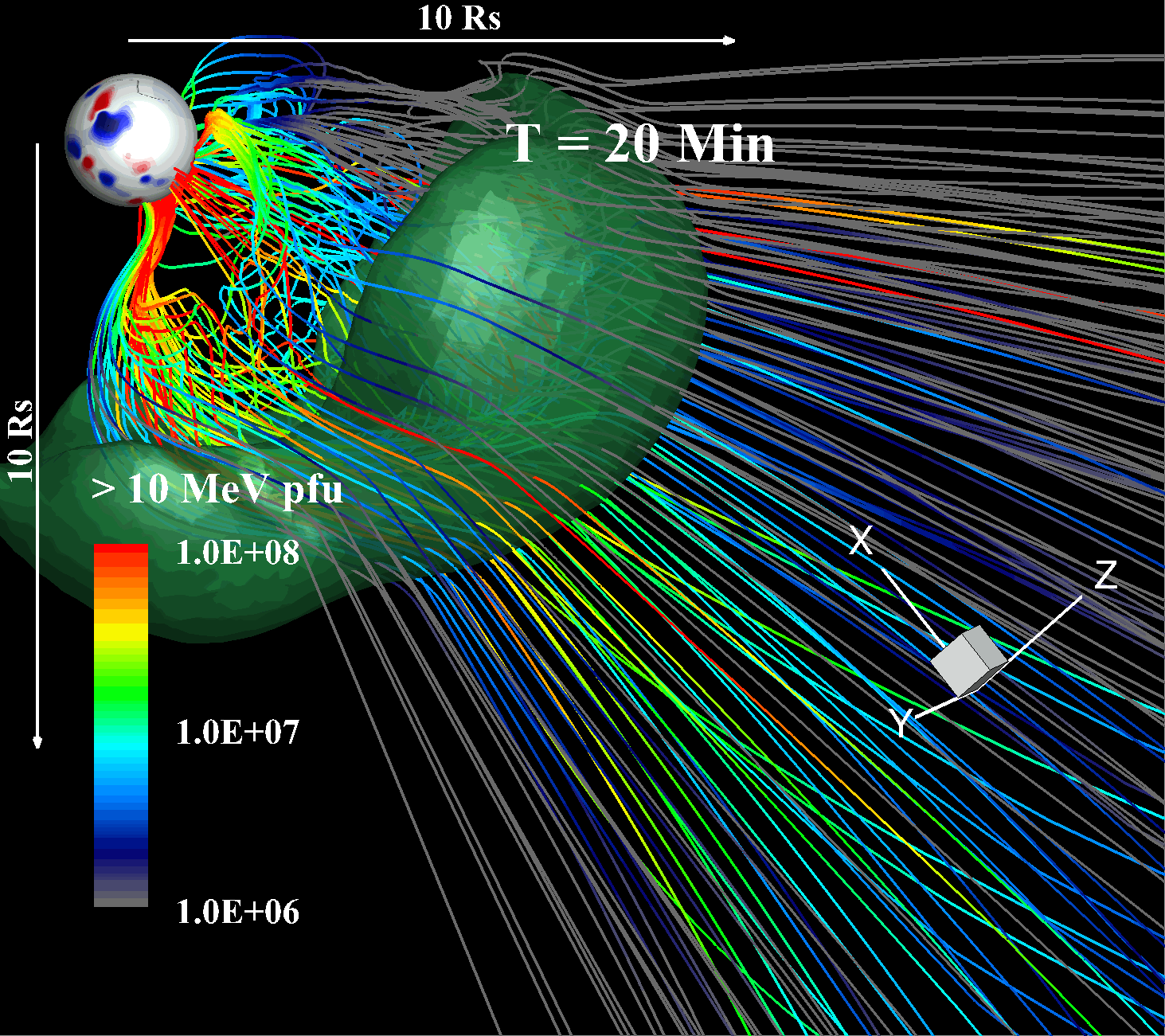}
    \caption{Distribution of $>\!\!10$ MeV protons along the extracted magnetic field lines at 20 min after the eruption of CME.
    The green isosurface represents the leading edge of the CME.
    }\label{fig:mflampa}
\end{figure}
M-FLAMPA, is developed at the University of Michigan \cite{Sokolov2004, Borovikov2018} and has been fully coupled with AWSoM(-R) and EEGGL in SWMF. 
M-FLAMPA extracts magnetic field lines from the solution of AWSoM(-R) and solves the Parker diffusion equation along a multitude of the extracted magnetic field lines.
Figure~\ref{fig:mflampa} illustrates hundreds of extracted magnetic field lines with M-FLAMPA during an SEP event. The sun is colored with the radial magnetic field and the green isosurface represents the leading edge of the CME. The distribution of the energetic particles ($>$ 10 MeV) along the extracted magnetic field lines at 20 min after the eruption of CME are shown in the unit of pfu. 

\subsubsection{AMPS}
\label{subsubsec:amps}

The Adaptive Mesh Particle Simulator (AMPS), also developed at the University of Michigan, is a 3-dimensional physics-based kinetic particle model for simulating the dynamics of neutral and charged particles.
AMPS incorporates two ways to model the transport of SEPs in the heliosphere: (1) simulating the transport of SEPs as they move along a set of evolving magnetic field lines extracted by M-FLAMPA, as demonstrated in \figurename~\ref{fig:mflampa}, and (2) simulating SEP transport in the 3D magnetic fields.
In previous applications, the model was successfully applied to study planetary, magnetospheric, and heliospheric environments \cite{Tenishev2021}. AMPS is a fully integrated component of SWMF.
AMPS solves the Parker and Focused transport equations in full 3D along a set of magnetic field lines \cite{Tenishev2005}.  AMPS compliments M-FLAMPA by sampling the particle distribution from a vast number of simulated particle trajectories, rather than calculating a distribution function. Combining these two approaches improves the chances of success. 

\subsubsection{iPATH/SEPCaster} 
\label{subsubsec:ipath}

IPATH, the improved Particle Acceleration and Transport in the Heliosphere model, developed at the University of Alabama in Huntsville, is a 2-D MHD-SEP model that simulates diffusive shock acceleration at CME-driven shocks and follows the subsequent transport of energetic particles in the ecliptic plane through the inner heliosphere. 
The iPATH model solves the transport of energetic particles by casting the focused transport equation into a set of stochastic differential equations. It is a Monte Carlo code following individual quasi-particles 
and is set up for parallel computations. 
Its recent extension, the SEPCaster model \cite{Li2021}, is a SEP model which couples the solar wind and CME model, AWSoM-R, with the particle acceleration and transport model iPATH. Instead of imposing an arbitrary inner boundary, the SEPCaster model follows the propagation of the CME and its driven shock from the corona base. 
In the Center, iPATH/SEPCaster will be coupled with the AWSoM(-R) and EEGGL/TiDeS-G modules.

\begin{figure}[bht]
    \vspace{-1.5ex}
    \includegraphics[width=1\textwidth]{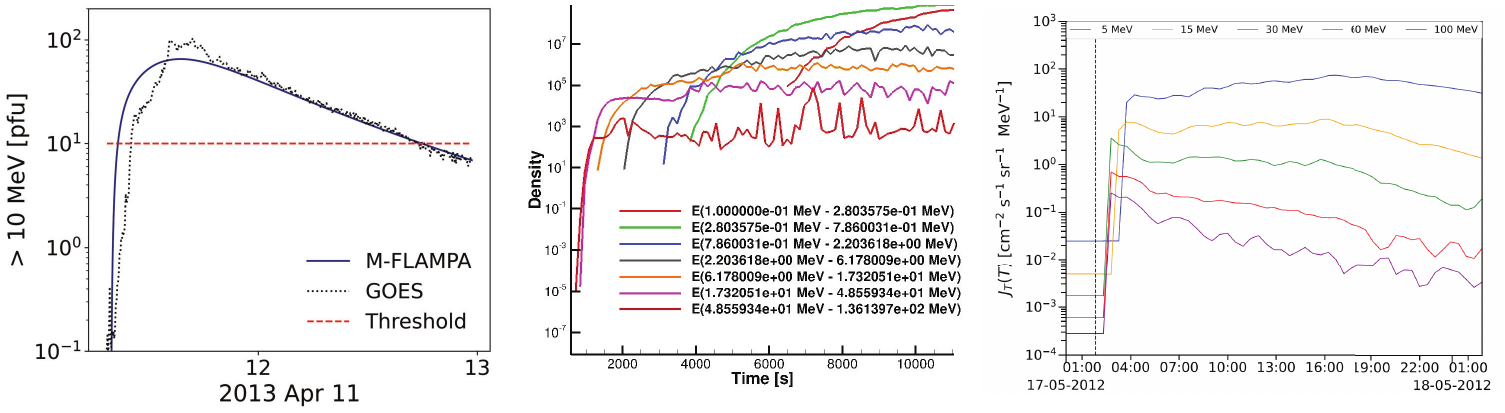}
    \caption{Left: $>10$ MeV proton flux comparison between M-FLAMPA results (blue solid) and GOES observations (dotted) for the 2013 Apr 11 SEP event. The red dashed horizontal line is the 10 pfu threshold. Middle: Particle density obtained from the AMPS model for the 2013 Apr 11 SEP event. Right: Differential proton intensity obtained from SEPCaster for the 2012 May 17 SEP event.
    }\label{fig:particle}
\end{figure}

Figure~\ref{fig:particle} shows the sample results of the three particle acceleration and transport modules, M-FLAMPA (left), AMPS (Middle), and iPATH/SEPCaster (right). 
In the CLEAR center, each particle module will be validated against historical events and their prediction results will also be evaluated collectively.
Both the boundary and initial conditions, input free parameters in different particle solver models will be validated and evaluated. 

\subsection{Seed Population Module}
\label{subsec:seed}

Diffusive shock acceleration requires a seed population of suprathermal particles that are energetic enough to cross the shock front multiple times \cite[and the references therein]{Zank2001} and get accelerated. The absolute particle flux also depends on the number of seed particles that are injected into the shock system \cite{Mewaldt2012,Hu2017,Li2021}. Therefore, the seed population is an important component in physics-based models. Three different seed particle injection profiles will be explored in the center:
(1) A suprathermal proton tail extending from the solar wind thermal distribution at the local kinetic temperature through to the injection energy, e.g., $\sim\!\!10$ keV \cite{Sokolov2004}. The local temperature and density are obtained from the AWSoM(-R) simulation;
(2) The injection profiles will be prescribed by scaling the suprathermal seed particle distribution observed at 1~AU back to close to the sun \cite{CCMC2022};
(3)  The third injection profile will be obtained by performing hybrid simulations to model the energization processes of (solar wind) thermal particles in collisionless shocks \cite{Karimo:2005,Caprioli:2015,Guo_Gia:2013, Caprioli:2014}.

We will perform 2D and 3D hybrid simulations of shock-driven ion acceleration with a state-of-the-art, multi-species, AI-powered code, HYPERS \cite{Omelchenko2012a}, the only hybrid code available currently for public use at the NASA CCMC. This 
massively parallel code implements a standard hybrid model, where ions are treated as full-orbit particles, the plasma electrons are modeled as an inertia-free quasineutral fluid, and radiation effects are neglected in Maxwell's equations \cite{Winske2003,Lipatov:2002}. 

Parameters that characterize the computational system, including the upstream flow Alfv{\'e}nic Mach number, $M_A$, measured in the downstream frame of reference, the obliquity angle, $\theta_{BN}$ between the background magnetic field and the direction of flow (shock normal), 
will be extracted from the AWSoM(-R) simulations. For a given set of shock parameters, hybrid simulations will provide the fraction of superthermal ions injected into the downstream 
together with their
pitch-angle information \cite{Caprioli:2015}. 
These distribution profiles will be input as seed population in the particle acceleration and transport modules.

\subsection{Validation and Readiness Levels of the CLEAR Tools}
In CLEAR, individual models and the integrated models will be validated using the benchmark SEP event dataset developed in this center. We will compare model outputs with observations and derive appropriate metrics for the parameters listed in the GAP analysis. 
Metrics for timing and fluxes will include mean error and mean log error, respectively, to measure bias; absolute mean error, absolute mean log error, and root mean square error to test accuracy. 
Metrics for the pre- and post-eruption ``All-Clear'' predictions will be measured in both the standard confusion matrix, skill scores (e.g., true skill score, Heidke skill score, BSS) and other derived skill scores in the Center. Validation for forecasting results is potentially subject to nonignorable variability when applied to data sets of limited sample size. The benchmark SEP dataset that we will work on falls within this realm and we are cautiously aware of this. We are planning on adopting principled statistical methodology in reliability, robustness and reproducibility \cite{yu2013stability, pineau2021improving} to properly quantify the uncertainty of our validation results and check the sensitivity of these. 
Model developers, observers, and theorists will develop metrics for evaluating the performance of the models on the prediction of pre- and post-eruption predictions and the time intensity profiles, including the availability of forecasts in a meaningful amount of time for end users, the occurrence of false alarms, and the accuracy of forecasts (e.g., peak intensity) when an SEP event does occur. 

The current technology readiness level (TRL) and readiness level (RL) of the two empirical models is TRL4/RL4: the system/model is validated in laboratory environment. The ML models are at the proof-of-concept level (TRL3/RL3). 
The physics-based models are between the TRL3/RL3 and TRL4/RL4. The proposed TRL/RL level of the integrated tool is TRL5/RL5: successful evaluation in the relevant environment. The validated system will be ready for forecaster/end user input at TRL5/RL5.

\section{Summary}
Improved SEP forecasting requires significant progress in several areas including: pre-eruption forecasting of flares and CMEs; accurate modeling of the background solar wind out to Jupiter's orbit and Interplanetary CME (ICME)/shock propagation; realistic CME generation; estimating the SEP seed population near the Sun and CME shocks; describing the magnetic connectivity between the Sun and any selected location in the heliosphere; particle acceleration processes, and particle transport.
Not only is the system large and varied, it is also strongly interconnected and coupled across multiple spatial and temporal scales. Only through the focused, multi-disciplinary approach to be followed by the CLEAR Center is it possible to ensure sufficient understanding of each of these areas and processes involved to allow progress to be made.

\section*{Open Research Section}
All the simulation data used in this paper including the 3D steady-state solution of the solar wind plasma and the time dependent flux profiles are publicly available at the Deep Blue Data Repository maintained by the University of Michigan.

\acknowledgments
This work was supported in part by NASA SWxC grant 80NSSC23M0191 (CLEAR). The author would like to acknowledge the entire CLEAR team.

\end{document}